\documentclass[11pt,a4]{article}

\def\slashchar#1{\setbox0=\hbox{$#1$}
   \dimen0=\wd0 \setbox1=\hbox{/} \dimen1=\wd1
   \ifdim\dimen0>\dimen1 \rlap{\hbox to \dimen0{\hfil/\hfil}} #1
   \else  \rlap{\hbox to \dimen1{\hfil$#1$\hfil}} / \fi}

\usepackage{graphicx}
\usepackage{cite}
\usepackage{amsmath}

\begin{document}

\title{Transversity structure of the pion in chiral quark models%
\thanks{Supported by the Bogoliubov-Infeld program (JINR), the Polish Ministry of
Science and Higher Education, grants N~N202~263438 and N~N202~249235,
Spanish DGI and FEDER grant FIS2008-01143/FIS, Junta de Andaluc{\'{\i}}a
grant FQM225-05, and EU Integrated Infrastructure Initiative Hadron Physics
Project, contract RII3-CT-2004-506078. AED acknowledges partial support from
the Russian Foundation for Basic Research, projects No.~10-02-00368 and No.~11-02-00112.}}

\author{{\bf W. Broniowski$^{1,2}$, E. Ruiz Arriola$^{3,4}$, A. E. Dorokhov$^{5,6}$ } \\~\\
$^1$Institute of Nuclear Physics PAN, PL-31342~Cracow, Poland\\
$^2$Institute of Physics, Jan Kochanowski University\\  PL-25406~Kielce, Poland\\ 
$^3$Departamento de F\'{\i}sica At\'omica, Molecular y Nuclear, Universidad de Granada \\E-18071 Granada, Spain\\
$^4$Instituto Carlos I de Fisica Te\'orica y Computacional, Universidad de Granada\\ E-18071 Granada, Spain\\
$^5$Joint Institute for Nuclear Research, Bogoliubov Laboratory of Theoretical Physics\\ RU-141980 Dubna, Russia \\
$^6$Institute for Theoretical Problems of Microphysics, Moscow State University\\ RU-119899 Moscow, Russia}

\date{Presented by WB at Mini-Workshop Bled 2011: \\Understanding Hadronic Spectra, Bled (Slovenia), 3-10 July 2011}

\maketitle

\begin{abstract}
We describe the chiral quark model evaluation of the 
transversity Generalized Parton Distributions (tGPDs) and related
transversity form factors (tFFs) of the pion. The obtained 
tGPDs satisfy all necessary formal requirements, such as the 
proper support, normalization, and
polynomiality. The lowest tFFs,  after the necessary QCD evolution, 
compare favorably to the recent lattice QCD 
determination. Thus the transversity observables of the pion  
support once again the fact that the spontaneously broken chiral
symmetry governs the structure of the Goldstone pion. The proper QCD 
evolution is crucial in these studies.
\end{abstract}


This talk is based on our two recent works
\cite{Broniowski:2010nt,Dorokhov:2011ew}, where more
details and a complete list of references may be found.  
Its topic concerns the transversity Generalized Parton Distribution (tGPD) of the pion, the least-known of the Generalized
Parton Distributions (see~\cite{Belitsky:2005qn,Feldmann:2007zz,Boffi:2007yc} and references
therein for an extensive review). 
The definition involves aligned parton-helicity operators (maximum-helicity case).
For the case of spin-0 hadrons, tGPDs arise because of the nonzero orbital angular
momentum between the initial and final state, thus allowing
to study the spin structure without the inherent complications of the explicit spin degrees of freedom, 
as in the case of the nucleon. In that situation the analysis of the spin structure of the pion is
particularly appealing, however, the quantity will be very difficult to access experimentally.

\begin{figure}[tb]
\centerline{\includegraphics[width=.57\textwidth]{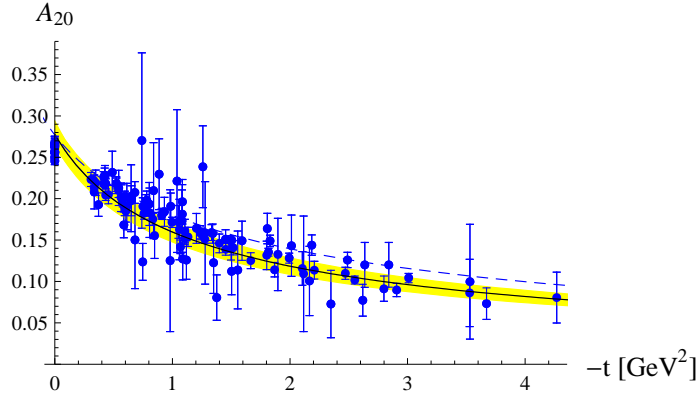}} 
\caption{The quark part of the spin-2 gravitational form factor in 
  Spectral Quark Model (solid line) and NJL model with the Pauli-Villars regularization (dashed line), compared to the lattice data from
  Ref.~\cite{Brommel:2005ee,Brommel:PhD}. The band around the Spectral Quark Model results corresponds
  to the model parameter uncertainty. \label{fig:gv}}
\end{figure}

A few years ago, however, lattice simulations \cite{Brommel:2007xd} provided
the lowest-order pion quark transversity form
factors (tFFs), defined as Mellin moments of tGPDs in the Bjorken $x$ variable. 
That way lattices supply valuable information on the nontrivial spin structure of hadrons.
In general, lattice calculations are capable to determine quantities that may only be dreamed off
to be measured experimentally and, in that regard, are extremely useful. The results can be 
used to verify various theoretical approaches and models in their rich spectrum of predictions. 
An example is the gravitational form factor of the pion. Its lattice determination \cite{Brommel:2005ee,Brommel:PhD}
agrees remarkably well with the evaluation in chiral quark models \cite{Broniowski:2008hx}, as can be seen from Fig.~\ref{fig:gv}.

Our study consist of two distinct parts: 1)~the chiral quark model determination
of tFFs and tGPDs of the pion and 2)~the QCD evolution.  For the first
part we apply the standard {\em local} NJL model with the
Pauli-Villars regularization \cite{RuizArriola:1991gc} and two
versions of the {\em nonlocal} models, where the quark mass depends on the momentum of the quark, namely, 
the instanton model \cite{Diakonov:1985eg} and
the Holdom-Terning-Verbeek (HTV) model \cite{Holdom:1990iq}.
We stress that chiral quark models have been successfully used for the evaluation of {\em soft}
matrix elements entering numerous high-energy processes
\cite{Davidson:1994uv,Davidson:2001cc,%
Broniowski:2003rp,Dorokhov:1998up,Polyakov:1999gs,Dorokhov:2000gu,Anikin:2000th,%
Praszalowicz:2003pr,Bzdak:2003qe,Nguyen:2011jy,Theussl:2002xp,Bissey:2003yr,%
Noguera:2005cc,Broniowski:2007si,Frederico:2009pj,Frederico:2009fk,Esaibegian:1989uj,Dorokhov:1991nj,%
Petrov:1998kg,Anikin:1999cx,Praszalowicz:2001wy,Dorokhov:2002iu,RuizArriola:2002bp,RuizArriola:2002wr,Broniowski:2008hx,%
Tiburzi:2005nj,Broniowski:2007fs,Courtoy:2008af,Kotko:2008gy}. They also agree with the Euclidean lattice determination of 
moments (see, e.g., \cite{Musch:2010ka,Hagler:2009ni}) 
and direct results from the transverse lattices \cite{Burkardt:2001jg,Dalley:2001gj,Dalley:2004rq,Dalley:2002nj}.

\begin{figure}[bt]
\centering
\includegraphics[width=0.33\textwidth]{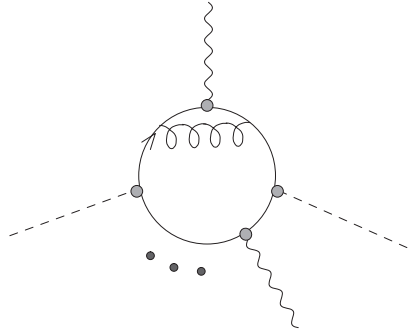}
\caption{One-loop (large-$N_c$) quark 
diagram, with external gauge bosons (wavy lines) and Goldstone mesons (dashed lines). The renorm-improved 
gluon exchanges are incorporated in terms of the LO DGLAP evolution. \label{fig:diagl}}
\end{figure}

The second element, crucial in obtaining proper results, is the QCD evolution, 
where renorm-improved radiative gluonic corrections are appended. 
The method is schematically depicted in Fig.~\ref{fig:diagl}. One-loop (large-$N_c$) quark 
diagram, with external gauge bosons and Goldstone mesons, is evaluated. Then the renorm-improved 
gluon exchanges are incorporated in terms of the LO DGLAP evolution.
The scale where the quark model calculation is
carried out can be identified with the help of the momentum fraction
carried by the quarks. According to phenomenology~\cite{Sutton:1991ay,Gluck:1999xe} or lattice
calculations~\cite{Best:1997qp}, the valence quark contribution is 47\%
of the total at the scale $\mu= 2{\rm GeV}$.  Since the quark
models possess no explicit gluons, the valence
quarks carry 100\% of the momentum. This determines the quark
model scale, denoted as $\mu_0$, as the scale determined in such a way, that 
when the evolution is carried out from $\mu_0$  up to $\mu=2$~GeV, the fraction
drops to $47\pm2$\%. The result of the LO DGLAP evolution is
\begin{eqnarray}
\mu_0=313^{+20}_{-10}~{\rm MeV}.  
\end{eqnarray}
Despite the low value of this scale, the prescription
has been successfully confirmed by a
variety of high-energy data and lattice calculations (see
\cite{Broniowski:2007si} and references therein). Moreover, the NLO
DGLAP modifications yield moderate corrections~\cite{Davidson:2001cc}, supporting 
our somewhat strained use of perturbative QCD at low scales. To summarize, 
{\em our approach = chiral quark model + QCD evolution}.

We now come to definitions. The pion $u$-quark tFFs, denoted as $B_{ni}^{\pi ,u}\left( t\right) $,
are defined via \cite{Diehl:2010ru}
\begin{eqnarray}
\langle \pi^{+}( p^{\prime }) | \overline{u}(0) i\sigma ^{\mu \nu }a_{\mu }b_{\nu }
\left( i \overleftrightarrow{D}a \right)^{n-1} u(0) | \pi^{+} (p) \rangle
=( a \cdot P )^{n-1} \frac{[ a \cdot p \, b \cdot p^{\prime}] }
{m_{\pi }}\sum_{\substack{ i=0,  \\ {\rm even}}}%
^{n-1}\left( 2\xi \right) ^{i}B_{ni}^{\pi ,u}\left( t\right) ,  \label{PionTme}
\end{eqnarray}
where the auxiliary vectors $a$ and $b$ satisfy the conditions $a^{2}=(ab)=0$ and $b^{2}\neq 0$. 
The skewness parameter is defined as $\xi =-{a \cdot q }/{( 2 a \cdot P ) }$,
the symbol $\overleftrightarrow{D}^{\beta }=\overleftrightarrow{\partial }^{\beta
}-igA^{\beta }$ is the covariant derivative, and $\overleftrightarrow{%
\partial }^{\beta }=\frac{1}{2}\left( \overrightarrow{\partial }^{\beta }-%
\overleftarrow{\partial }^{\beta }\right) $. Next, $p^{\prime }$
and $p$ are the initial and final pion momenta, $P=\frac{1}{2}(p^{\prime
}+p) $, $q=p^{\prime }-p$, and $t=-q^{2}$. 
The bracket denotes antisymmetrization in the vectors $a$
and $b$. The corresponding definition of the tGPD is \cite{Belitsky:2005qn}
\begin{eqnarray}
\langle \pi ^{+}(p^{\prime })\mid \bar{u}(-a)i\sigma ^{\mu \nu }a_{\mu}b_{\nu }u(a)\mid \pi ^{+}(p)\rangle
=\frac{\left[  a \cdot p\,  b \cdot p^{\prime } \right] }
{m_{\pi }}\int_{-1}^{1}dx \,e^{-i x\, P \cdot a }E_{T}^{\pi, u}(x,\xi ,t), \label{PionTGPD}
\end{eqnarray}
where the presence of the gauge link operator are understood. 
The tFFs for the $d$-quarks follow from the isospin symmetry, namely 
$B_{ni}^{\pi ,d}\left( t\right) =\left( -1\right) ^{n}B_{ni}^{\pi ,u}\left( t\right)$. The tFFs are the 
moments of the tGPD in the $x$-variable,
\begin{equation}
\int_{-1}^{1}dx\,x^{n-1}E_{T}^{\pi, u}\left( x,\xi ,t\right)
=\sum_{\substack{ i=0, \\ {\rm even}}}^{n-1}\left( 2\xi \right)^{i}B_{ni}^{\pi ,u}\left( t\right).  \label{En}
\end{equation}
This formula explicitly displays the desired polynomiality property. We remark that the full information carried
by tGPDs is contained in the collection of the infinitely many tFFs. 


\begin{figure}[tb]
\centering
\includegraphics[width=0.49\textwidth]{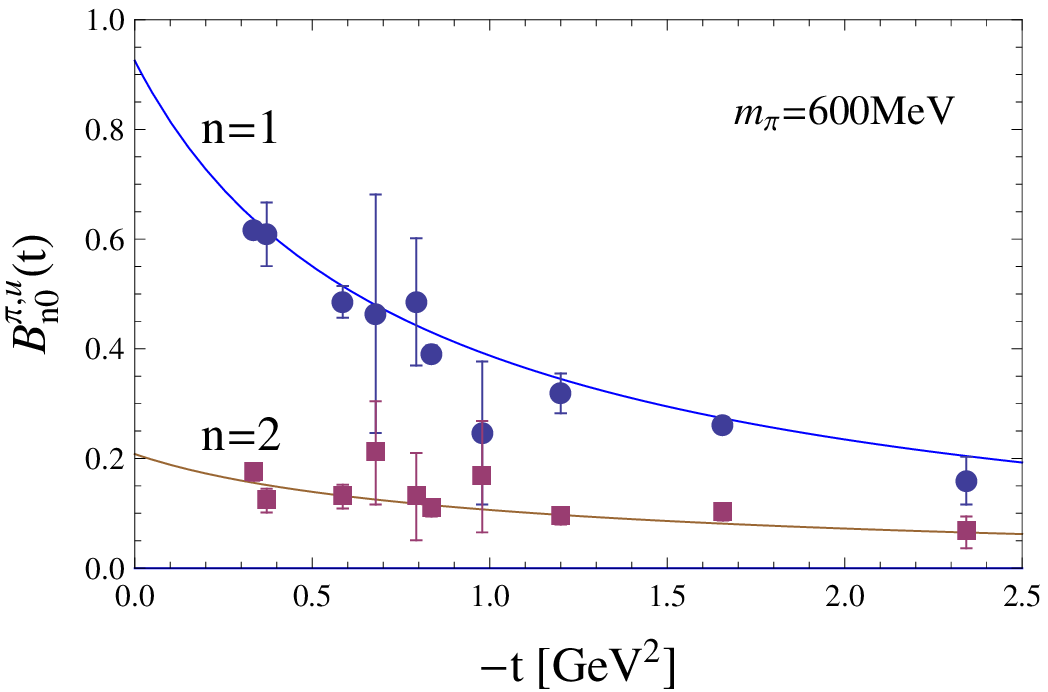} \includegraphics[width=0.49\textwidth]{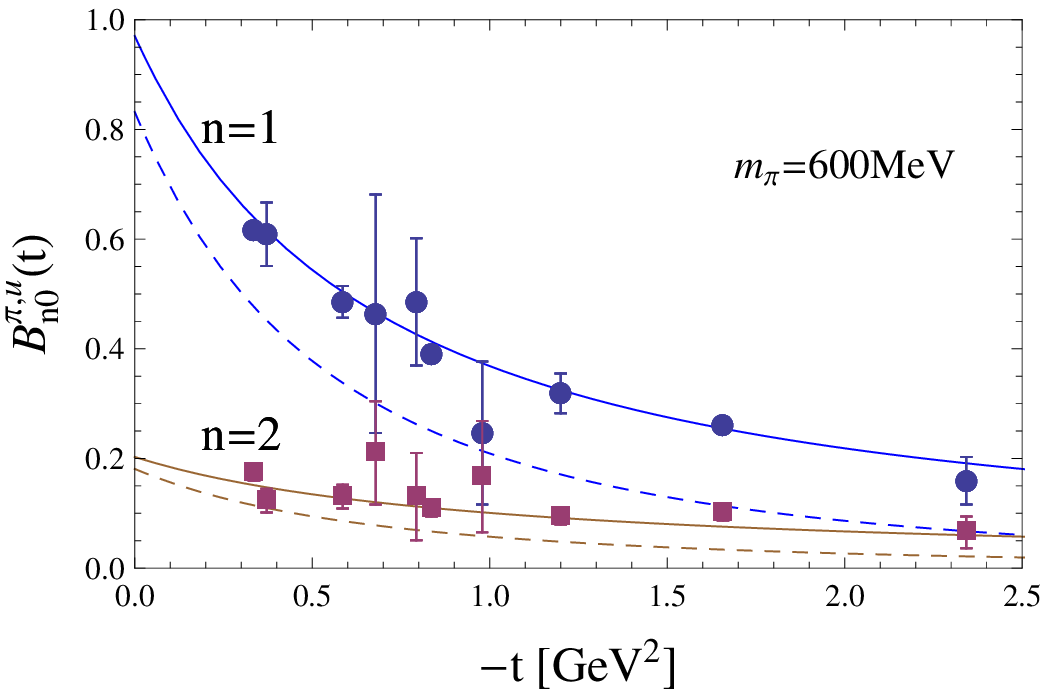}
\caption{The form factors $B^{\pi,u}_{10}(t)$ and $B^{\pi,u}_{20}(t)$, evaluated at $m_\pi=600$MeV in the local
NJL model (left panel) and in nonlocal models (right panel, solid line -- HTV model, dashed line -- instanton model).
The lattice data from~\cite{Brommel:2007xd}. The local NJL and HTV models agree very well with the data. \label{fig:B12}}
\end{figure}

The full details of the quark-model calculation as well as the QCD evolution can be found in~\cite{Broniowski:2010nt,Dorokhov:2011ew}. The
two lowest tFFs available from the lattice data, $B_{10}^{\pi,u}$ and $B_{20}^{\pi ,u}$, evolve multiplicatively
in a simple way:
\begin{equation}
B_{n0}^{\pi ,u}(t;\mu )=B_{n0}^{\pi ,u}(t;\mu _{0})\left( \frac{\alpha
(\mu )}{\alpha (\mu _{0})}\right) ^{\gamma _{n}^{T}/(2\beta _{0})},
\end{equation}%
where $\gamma _{n}^{T}$ are the appropriate anomalous dimensions~\cite{Broniowski:2010nt,Dorokhov:2011ew}.
In the local model, in the chiral limit at $t=0$ we find the very simple result:
\begin{eqnarray}
B_{10}^{\pi,u}(t=0;\mu_{0})/m_{\pi}=\frac{N_{c} M}{4\pi^{2} f_{\pi}^{2}}, \; \;\;
\frac{B_{20}^{\pi,u}(t=0;\mu)}{B_{10}^{\pi,u}(t=0;\mu)}=\frac{1}{3}
\left( \frac{\alpha(\mu)}{\alpha(\mu_{0})}\right) ^{8/27},  \label{LocLim2}
\end{eqnarray}
where $M$ is the constituent quark mass.
The results of the model calculation followed by evolution are shown in Fig.~\ref{fig:B12}. We note a striking agreement with the lattice data
\cite{Brommel:2007xd} for the local NJL model, as well as for the non-local HTV model. 

\begin{figure}[tb]
\centering
\includegraphics[width=0.49\textwidth]{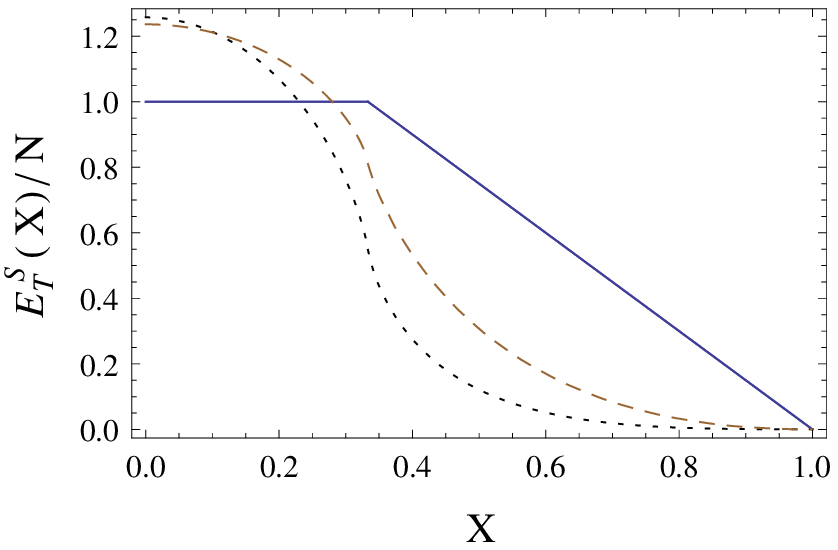} \includegraphics[width=0.49\textwidth]{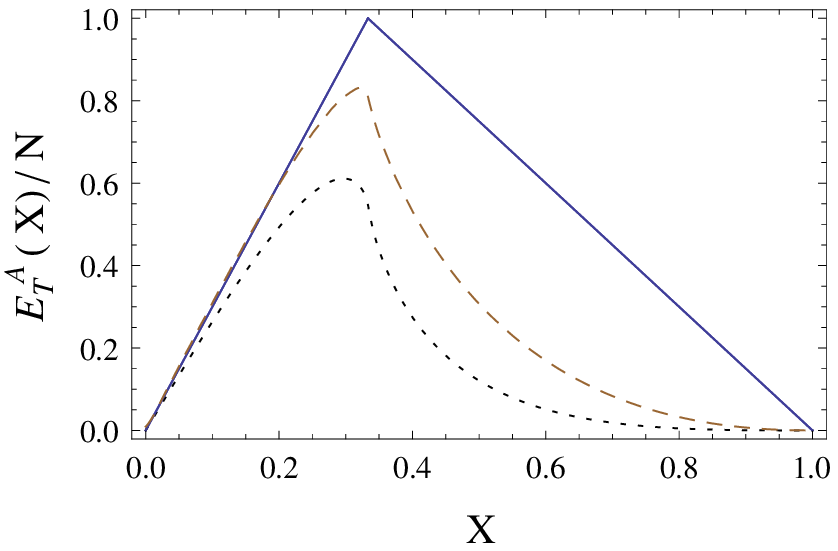}
\caption{The symmetric (left panel) and antisymmetric (right panel) tGPDs of the pion at $t=0$ and $\xi=1/3$, evaluated in the
NJL model in the chiral limit at the quark-model scale $\mu_0=313$~MeV (solid lines) and  evolved to the
scales $\mu=2$~GeV (dashed lines) and $1$~TeV (dotted lines). \label{fig:tGPD}}
\end{figure}

\begin{figure}[tb]
\centering
\includegraphics[width=0.49\textwidth]{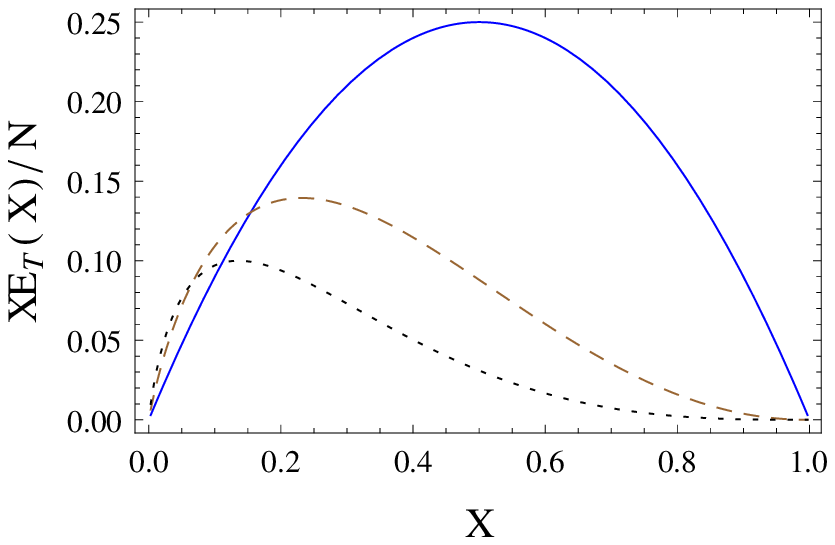} \includegraphics[width=0.49\textwidth]{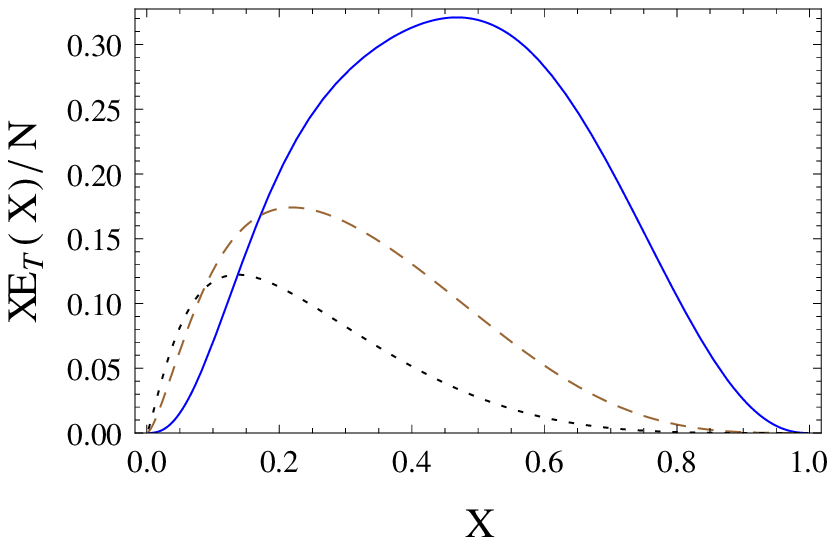}
\caption{The tGPD of the pion at $t=0$ and $\xi=0$, evaluated in the chiral limit in the
local NJL model (left panel) and in the instanton model (right panel). The solid lines
correspond to the quark-model scale $\mu_0=313$~MeV, the dashed lines to $\mu=2$~GeV, 
and the dotted lines to $\mu=1$~TeV. \label{fig:locnloc}}
\end{figure}

Finally, we present the results for the full tGPD for $t=0$ and $\xi = 1/3$ or $\xi=0$. 
The evolution is different for the symmetric and antisymmetric parts of tGPDs, 
hence we define the isovector and isoscalar combinations:
\begin{eqnarray}
&& E_{T}^{\pi ,I=1}\left( x,\xi ,Q^{2}\right) \equiv E_{T}^{\pi ,S}\left( x,\xi ,Q^{2}\right) = E_{T}^{\pi }\left(
x,\xi ,Q^{2}\right) +E_{T}^{\pi }\left( -x,\xi,Q^{2}\right) ,   \nonumber \\
&& E_{T}^{\pi ,I=0}\left( x,\xi ,Q^{2}\right) \equiv E_{T}^{\pi ,A}\left( x,\xi ,Q^{2}\right) \nonumber = E_{T}^{\pi }\left(
x,\xi ,Q^{2}\right) -E_{T}^{\pi }\left( -x,\xi,Q^{2}\right) .  \label{ETnI01}
\end{eqnarray}
The QCD evolution has been carried out with the method of~\cite%
{Kivel:1999sk,Kivel:1999wa,Manashov:2005xp,Kirch:2005tt}. The results for $\xi=1/3$ in the NJL model are shown
in Fig.~\ref{fig:tGPD}, while in Fig.~\ref{fig:locnloc} we compare the
result for $\xi=0$ in the NJL model (left panel) and the nonlocal instanton model (right panel). Except for different
end-point behavior, discussed in~\cite{Dorokhov:2011ew}, the results are similar.

In conclusion we wish to stress that the
absolute predictions for the multiplicatively evolved $B_{10}$ and $B_{20}$ agree
remarkably well with the lattice results, supporting the assumptions of numerous other calculations
following the same ``chiral quark model + QCD evolution'' scheme.
Our study of the transversity observables of the pion  
support once again the feature that the spontaneously broken chiral
symmetry determines the structure of the Goldstone pion.


\end{document}